\documentclass[12pt,preprint]{aastex}

\usepackage{epsfig,graphicx,amssymb,epstopdf,mathrsfs}

\newcommand{\rxsj}{1RXS~J141256.0+792204}
\newcommand{\calvera}{Calvera}
\newcommand{\eighteen}{1RXS~J185635.1$-$375433}
\newcommand{\objtwo}{CXOU~J141252.46+792152.6}

\newcommand{\swift}{\textit{Swift}}

\newcommand{\chandra}{\textit{Chandra}}

\newcommand{\xmm}{\textit{XMM-Newton}}

\def\nh{\mbox{$N_{\rm H}$}}

\def\deg{\hbox{$^\circ$}}

\def\arcmin{\hbox{$^\prime$}}
\def\arcsec{\hbox{$^{\prime\prime}$}}
\def\ctsec{\hbox{${\rm cts}\,{\rm s^{-1}}$}}
\def\ctsframe{\hbox{${\rm cts}\,{\rm frame^{-1}}$}}
\def\ctsksec{\hbox{${\rm cts}\,{\rm ks^{-1}}$}}
\def\kmsec{\hbox{${\rm km}\,{\rm s^{-1}}$}}
\def\km{\hbox{${\rm km}$}}
\def\simlt{\mathrel{\hbox{\rlap{\hbox{\lower4pt\hbox{$\sim$}}}\hbox{$<$}}}}
\def\simgt{\mathrel{\hbox{\rlap{\hbox{\lower4pt\hbox{$\sim$}}}\hbox{$>$}}}}
\def\chisqr{\mbox{$\chi^2$}}
\def\chisqrnu{\mbox{$\chi^2_\nu$}}

\newcommand{\Msun}{\mbox{$M_\odot$}}

\newcommand{\xray}{\mbox{X-ray}}
\newcommand{\etal}{\mbox{et al.}}

\newcommand{\percmsq}{cm$^{-2}$}

\newcommand{\perval}[2]{{#1\mbox{$^{#2}$}}}

\newcommand{\persec}{\perval{\rm s}{-1}\/}
\newcommand{\percm}{\mbox{$\cm^{-2}$}}

\newcommand{\ppm}{\mbox{$\pm$}}

\newcommand{\tee}[1]{\mbox{$\times 10^{#1}$}}

\newcommand{\erg}{\mbox{$\rm\,erg$}\/}
\newcommand{\cm}{\mbox{$\rm\,cm$}}

\newcommand{\kteff}{\mbox{$kT_{\rm eff}$}}

\newcommand{\kTeffinfty}{\mbox{$kT_{\rm eff}^\infty$}}
\newcommand{\nyqnu}{\mbox{$\nu_{\rm Nyq}$}}

\newcommand{\ud}[2]{\mbox{$^{+ #1}_{- #2}$}}
\newcommand{\cgsflux}{\erg\,\percm\,\persec}

\begin{document}

\title{Chandra Observations of \rxsj\ (\calvera)}

\author{A.~S.~H. Shevchuk\altaffilmark{1},
        D.~B. Fox\altaffilmark{1}, \&
        R.~E. Rutledge\altaffilmark{2}}

\email{ahs148@psu.edu, dfox@astro.psu.edu, rutledge@physics.mcgill.ca}

\altaffiltext{1}{Department of Astronomy \& Astrophysics, 525 Davey
  Laboratory, Pennsylvania State University, University Park, PA
  16802, USA} 

\altaffiltext{2}{Department of Physics, McGill University, 3600 rue
  University, Montreal, QC H3A 2T8, Canada}


\shorttitle{Chandra Observations of Calvera}
\shortauthors{Shevchuk, Fox \& Rutledge}


\begin{abstract}
We report the results of a 30\,ks \chandra/ACIS-S observation of the
isolated compact object \rxsj\ (\calvera).  The \xray\ spectrum is
adequately described by an absorbed neutron star hydrogen atmosphere
model with $\kTeffinfty = 88.3\pm 0.8$\,eV and radiation radius
$R^\infty/d = 4.1\pm 0.1$\,km~kpc$^{-1}$.  The best-fit blackbody
spectrum yields parameters consistent with previous measurements;
although the fit itself is not statistically acceptable, systematic
uncertainties in the pile-up correction may contribute to this.
We find marginal evidence for narrow spectral features in the
\xray\ spectrum between 0.3 and 1.0 keV.  In one interpretation, we
find evidence at 81\%-confidence for an absorption edge at
$E=0.64\ud{0.08}{0.06}$\,keV with equivalent width ${\rm EW}\approx
70$\,eV; if this feature is real, it is reminiscent of features seen
in the isolated neutron stars RX~J1605.3+3249, RX~J0720.4$-$3125, and
1RXS~J130848.6+212708 (RBS~1223). 
In an alternative approach, we find evidence at 88\% -confidence
for an unresolved emission line at energy $E=0.53\pm 0.02$\,keV, with
equivalent width ${\rm EW}\approx 28$\,eV; the interpretation of this
feature, if real, is uncertain.  
We search for coherent pulsations up to the Nyquist frequency
$\nyqnu=1.13$\,Hz and set an upper limit of 8.0\% rms on the strength
of any such modulation.  We derive an improved position for the source
and set the most rigorous limits to-date on any associated extended
emission on arcsecond scales.  Our analysis confirms the basic picture
of \calvera\ as the first isolated compact object in the ROSAT/Bright
Source Catalog discovered in six years, the hottest such object known,
and an intriguing target for multiwavelength study.
\end{abstract}

\keywords{X-rays: stars --- 
          stars: neutron --- \\ \hspace*{0.5in}
          X-rays: individual: \rxsj\ ---
          methods: statistical}

\maketitle


\section{Introduction}
\label{sec:intro}

Between 1996 and 2001, seven of the 18,811 sources in the ROSAT Bright
Source Catalog (BSC; \citealt{voges99}) were identified as radio-quiet
neutron stars without associated supernova remnants or binary
companions. These seven isolated neutron stars (INSs;
\citealt{haberl05}) have the following properties: \xray-to-optical
flux ratios exceeding 1000, thermal spectra peaking in the far-UV or
soft \xray, minimal ($<$10\%) long-term \xray\ variability, and
rotation periods of several seconds.  These properties are consistent
with interpretation of the INSs as a population of $\sim$Myr-old,
cooling, non-accretion-powered neutron stars.

The spectra of INSs show little evidence of non-thermal emission, with
atmospheres that are thought to be well-suited for theoretical
modeling \citep{rajagopal96,pavlov96} that may ultimately lead to
constraints on neutron star physical parameters and the equation of
state (EOS; \citealt{lattimer04,page06}). However, the true EOS cannot
be distinguished from among the plethora of possible EOSs consistent
with current theories of quantum chromodynamics without multiple data
points \citep{vankerkwijk03}, and a sample of seven is likely too
small to effectively solve this puzzle.

For more than six years, and despite substantial efforts (e.g.,
\citealt{rutledge03b,agueros06}), no new INSs were discovered in the
BSC.  In 2005 we initiated a new approach to this problem, identifying
BSC sources likely to have high \xray-to-optical flux ratios and using
NASA's \swift\ satellite \citep{swift} to efficiently survey these
targets (see \citealt{fox04}; Shevchuk \etal\ in prep; Letcavage
\etal\ in prep). Based on the results of our survey effort, now
substantially complete, \citet{turner09} estimate that the BSC
contains fewer than 48 INSs at 90\% confidence, assuming all-sky
isotropy. As such, the BSC in total may provide only a modest increase
in the number of known INSs; nonetheless, it remains a promising
catalog to explore for these objects.  At fainter flux levels, a
strong candidate INS, 2XMM~J104608.7$-$594306, has recently been
discovered with \xmm, and discovery of more such objects may be
anticipated in the future \citep{pires09}.

The first confirmed isolated compact object (ICO) from our survey of
BSC sources is \rxsj\ (\calvera).  Initial observations of
\calvera\ left substantial room for interpretation
\citep{rutledge08}. \calvera's temperature, as determined by a
blackbody model of the \swift\ spectrum, is $215\pm{25}$\,eV,
approximately twice that of 1RXS~J130848.6+212708, the hottest
previously-known INS. Combined with its Galactic latitude of 37\deg,
\calvera\ must be $\approx$5\,kpc above the Galactic plane to conform
to the INS model. Current cooling models suggest it is not feasible
for a neutron star to move so far from its (likely) origin in the
Galactic plane while remaining so hot, which casts doubt on \calvera's
interpretation as an INS. Searching for a way to frame the
\swift\ results in the context of known neutron star classes, it was
suggested that \calvera\ might be a millisecond radio pulsar (MSP) at
a distance of 80--260\,pc from the Sun \citep{rutledge08}. This claim
was based on an exploration of a range of possible distances and
(corresponding) \xray\ luminosities for \calvera, which -- for
reasonable distance values -- yielded consistent results only in
comparison to the properties of the 47~Tuc MSP population.  Subsequent
radio observations yielded no evidence of radio pulsations to deep
luminosity limits \citep{hessels07}, suggesting that if \calvera\ is a
nearby radio MSP, it is beamed away from Earth.

The aim of the present work is to examine \calvera's \xray\ properties
via our recent 30\,ks \chandra\ ACIS-S observation, the most detailed
investigation of the source to-date. In \S\ref{sec:obs} we discuss
the parameters of this observation, our analysis, and the models used
to interpret the data, addressing in turn the astrometric, spatial,
timing, and spectroscopic implications of the \chandra\ data.
\S\ref{sec:discuss} discusses these results in the larger context of
INS studies, and \S\ref{sec:conclude} summarizes our conclusions.


\section{Observations \& Analysis}
\label{sec:obs}

Data were analyzed using CIAO v4.0.1 \citep{ciao} with
\chandra\ calibration database CALDB v3.4.4.

\chandra\ observed \rxsj\ with the ACIS-S detector beginning 8~Apr
2008 03:42:08~TT, in a single pointing.  The standard 1/8 subarray was
used, giving a time resolution of 0.44104\,s, with the target placed
at the center of the subarray window and offset from the boundary
between readout quadrants, according to usual practice.  Based on the
observed source count rate ($\sim$0.2~\ctsec), and accounting for
9.3\% deadtime per frame, we observe 0.080~\ctsframe.  Given this
per-frame count rate, a mild level of pile-up is anticipated (see
\S\ref{sub:obs:spectral}), which will affect the source spectrum and
PSF profile.

The total observation time was 29,142\,s; due to CCD read-out, the
total exposure time for the observation was 9.3\% less, with 26,430\,s
of exposure (indeed, a small number of ``out of time'' events are
observed in the data).  The focal plane temperature was 153.28\,K
($-119.87\deg$\,C).

Using {\tt wavdetect} we detect \rxsj\ with a source detection
significance of 72$\sigma$.


\subsection{Astrometry}
\label{sub:obs:astrom}

Using uncorrected \chandra\ astrometry, we localize \calvera\ to
R.A. 14h12m55.84s, Dec.\ +79d22m03.75s (J2000) with 0.6\arcsec\ radius
(90\% confidence) uncertainty.  This is within the 90\% confidence
region of the previously-derived \xray\ position \citep{rutledge08}.

Using previously-analyzed Gemini-North + GMOS imaging of this region
\citep{rutledge08}, we identify optical counterparts to two
\xray\ sources in the \chandra\ field of view,
CXOU~J141220.78+792251.6 and CXOU~J141246.23+792222.3.  The two
registration sources provide a weighted average correction to the
\chandra\ native coordinates of $\Delta\alpha=-0.21\ppm0.14\arcsec$,
$\Delta\delta=-0.33\ppm0.12\arcsec$, where 1$\sigma$ uncertainties in
the correction are dominated by \chandra\ centroid uncertainties for
these faint sources.  Applying this correction yields our best
estimate for the position of \calvera: R.A.=14h12m55.76s,
Dec.=+79:22:03.4 (J2000) with 90\%-confidence ellipse semimajor axes
of 0.31\arcsec\ (R.A.) $\times$ 0.25\arcsec\ (Dec.),
aligned with the coordinate axes.

The brightest \xray\ source apart from \calvera\ is a 5.3$\sigma$ {\tt
  wavdetect} detection at R.A. 14h12m52.46s, Dec.\ +79d21m52.6s
(J2000), which has a statistical localization uncertainty of
0.1\arcsec.  This source does not have an optical counterpart in our
Gemini imaging; however, its relative brightness would allow it to be
used in an \xray\ proper motion study.  The offset between
\calvera\ and \objtwo\ is 14.5\ppm0.2\arcsec, which would allow a
(3$\sigma$) detection of relative proper motion between the two
sources for relative on-sky motions of
$$
v_{\rm sky} > 290 \left(\frac{d_{\rm 100\, pc}}{T_{\rm 1\, yr}}\right)
   \kmsec,
$$
where $d_{\rm 100\, pc}$ is the distance to \calvera\ in units of
100~pc and $T_{\rm 1 \, yr}$ is the elapsed time to the second-epoch
observation, in years.  This limiting velocity corresponds to $v_{\rm
  calvera}\approx 32,000$\,\kmsec\ at a distance of 11~kpc, the
distance implied for an INS interpretation of \calvera.

\subsection{Spatial Analysis}

In order to compare the observed spatial distribution of detected
\xray\ photons to point-spread-function (PSF) models, we used the
Chandra Ray Tracer (ChaRT) with MARX v4.3 \citep{chart} to produce a
ray-trace onto the detector plane using our best-fit \xray\ spectrum
over the 0.3--8.0\,keV photon energy range, using the default internal
model for dither-blur and a dither-blur radius of 0.35\arcsec.

Comparison of the radial count distributions of the observed and
simulated (MARX) datasets reveals that the observed distribution is
significantly \textit{narrower} than the MARX distribution, such that
the simulated PSF does not provide a statistically acceptable
description of the observed PSF, as evaluated with a two-sample K-S
test \citep{press}.  We interpret this as implying that the
\xray\ source remains unresolved at \chandra\ spatial resolution, to
within existing \chandra\ PSF modeling capabilities.

To derive limits on the fractional flux in any resolved component, we
simulated added Gaussian spatial components with widths of
$\sigma$=1\arcsec, 5\arcsec, and 10\arcsec, respectively.  For each
specified width, we add counts to the source image until the radial
distribution of counts is inconsistent with (and broader than) the
observed radial distribution of counts, at 90\% confidence, according
to a K-S test.  In this manner, we produce 90\%-confidence upper
limits on the fractional flux of any spatially-resolved component.
These limits are: $<$5\% for $\sigma=1\arcsec$, and $<$3\% for
$\sigma=5\arcsec$ and $\sigma=10\arcsec$.


\subsection{Timing Analysis}
\label{sub:obs:timing}

We extracted counts within a circle 5\arcsec\ of \calvera, and use a
300~pixel by 70~pixel off-source region to sample background counts,
finding 3761 total counts.  In the source region for \calvera\ we
detect 4958 counts, giving a background-subtracted count rate of
185\ppm3\,\ctsksec. The background represents 1.1\% of detected counts
and is neglected in the following analysis.

We barycenter-corrected the data using {\tt axbary}. Using counts at
energies $<$5\,keV, we produce a power density spectrum with
frequencies between 3.4\tee{-5}~Hz and 1.13~Hz, finding no evidence
for periodicity in these data, with a maximum Leahy-normalized power
\citep{leahy83} of 20.2.  We set a 90\%-confidence upper-limit
\citep{vaughan94} on the root-mean-square variability for a sinusoidal
signal in this frequency range:
$$
{\rm rms} < \sqrt{\frac{P}{N_{\rm phot}}} \frac{I}{I-B},
$$
where $P$ is the upper limit on the detected power, $N_{\rm phot}$ is
the number of photons, and $I$ and $B$ are the relative source and
background count rates, respectively.  Our limit corresponds to ${\rm
  rms} < 8.0\%$ variability.  Given that \eighteen\ is known to have
pulsations at the 1.2\%~rms level \citep{tiengo07}, this may not be
considered a strong constraint.

We performed the same analysis for all counts with energies $<$1\,keV
(2994 counts), and detect no periodicity, with a maximum power of
18.4; the rms upper-limit in this energy range is ${\rm rms} < 9.0\%$
($<$1\,keV). 


\subsection{Spectral Analysis}
\label{sub:obs:spectral}

Spectral analysis was performed with XSPEC v12.3.1x
\citep{xspec}. Working from the {\tt psextract} science thread on the
Chandra \xray\ Center website\footnote{http://cxc.harvard.edu}, we
used {\tt psextract} with a source aperture of 3.7\arcsec\ to extract
4,711 source region counts and two circular background regions on
opposing sides of the source, each with radius 30\arcsec, to extract a
total of 158 background counts. The charge bleed upon readout present
around the source prevented us from using an annular extraction
region.  We used {\tt mkacisrmf} to generate the response matrix and
{\tt mkarf} was then used to generate the ancillary response file. We
grouped the spectrum into bins with a minimum of fifty counts per bin
over the range 0.32 to 7.33\,keV (the maximum detected photon energy).

Using the PIMMS software
tool\footnote{http://heasarc.nasa.gov/Tools/w3pimms.html} we estimate
a pile-up fraction of 3\% for this observation. We account for this
level of pile-up in our analyses using the {\tt pileup} model in XSPEC
\citep{davis01}, with pile-up model values frozen at values
recommended in the \chandra\ ABC Guide to
Pileup\footnote{http://cxc.harvard.edu/ciao/download/doc/pileup\_abc.ps}:
the per-frame integration time is 0.40~s (appropriate to our
1/8-subarray dataset),
a maximum of five piled-up photons are modeled, 
the grade correction parameter {\tt g0} is unity,
the grade migration parameter $\alpha=0.5$, 
the PSF fraction is 0.95, 
and the number of regions used is one. The PSF fraction parameter
should equal the fraction of all extracted counts that land in the
central 3$\times$3 pixel island, which we find is an appropriate
estimate for \calvera; the meaning of the other model parameters is
discussed in detail in \citet{davis01}.  As discussed below, we find
that this pile-up model can account for the high-energy portion of the
observed data for reasonable models of the underlying source spectrum.

All uncertainties below are quoted at 90\%-confidence unless otherwise
specified. 


{\bf Power-Law.}
We find that a power-law spectrum does not provide an adequate fit to
the data ($\chisqrnu=2.31$ for $\nu=67$ degrees of freedom; $p =
7.3 \times 10^{-9}$). Incorporating pile-up corrections and allowing
the interstellar absorption column to vary freely, we find a best-fit
photon index $\alpha=4.8$ with $\nh = 3.4\times 10^{21}$\,\percmsq, in
excess of the total Galactic column along the line of sight ($\nh =
2.65\times 10^{20}$\,\percmsq\ from \ion{H}{1}\ maps;
\citealt{kalberla05}).  


{\bf Blackbody.}
We find that a blackbody spectrum does not fit the data acceptably
($\chisqrnu=2.04$ for $\nu=67$ degrees of freedom, $p = 1.0 \times 10^{-6}$). The best-fit
temperature for a blackbody with pile-up corrections and without
interstellar absorption (the equivalent neutral hydrogen column
density \nh, a free parameter, is driven to zero during the fit) is
$kT=229$\,eV, with blackbody normalization $(R_{\rm km}/D_{\rm
  10\,kpc})^2=26.6$. This fit is consistent with our original
\swift\ XRT observation, which yielded $kT = 215\pm 25$\,eV
\citep{rutledge08}.  For reference purposes only, we present the
parameters of our best-fit blackbody (without uncertainties, as these
are not well defined in the absence of an acceptable fit) in
Table~\ref{tab:params}.


{\bf Neutron Star Hydrogen Atmosphere.}
Fitting the spectrum of a neutron star hydrogen atmosphere model ({\tt
  nsa} model component in XSPEC; \citealt{zavlin96}), including
interstellar absorption and pile-up corrections, yields a fit with
$\chisqrnu=1.305$ ($\nu=67$) and best-fit parameters: effective
temperature $kT_{\rm eff}=109\pm 1$\,eV, equivalent neutral hydrogen
column density $N_H = 3.1\pm 0.9 \times 10^{20}$\,\percmsq, and
normalization $D_{\rm kpc}^{-2}=0.077_{-0.038}^{+0.041}$.  In using
the {\tt nsa} model we fix the neutron star mass at 1.4\,\Msun, its
radius at 12\,\km, and its magnetic field at zero; since this neutron
star model gives a gravitational redshift of $z_g = 0.235$, the
corresponding parameters for distant observers are $\kTeffinfty =
88.3\pm 0.8$\,eV, $R^\infty = 14.8$\,km, and $R^\infty/d = 4.1\pm
0.1$\,km~kpc$^{-1}$.

\begin{figure}
\centerline{\includegraphics[width=7.5in]{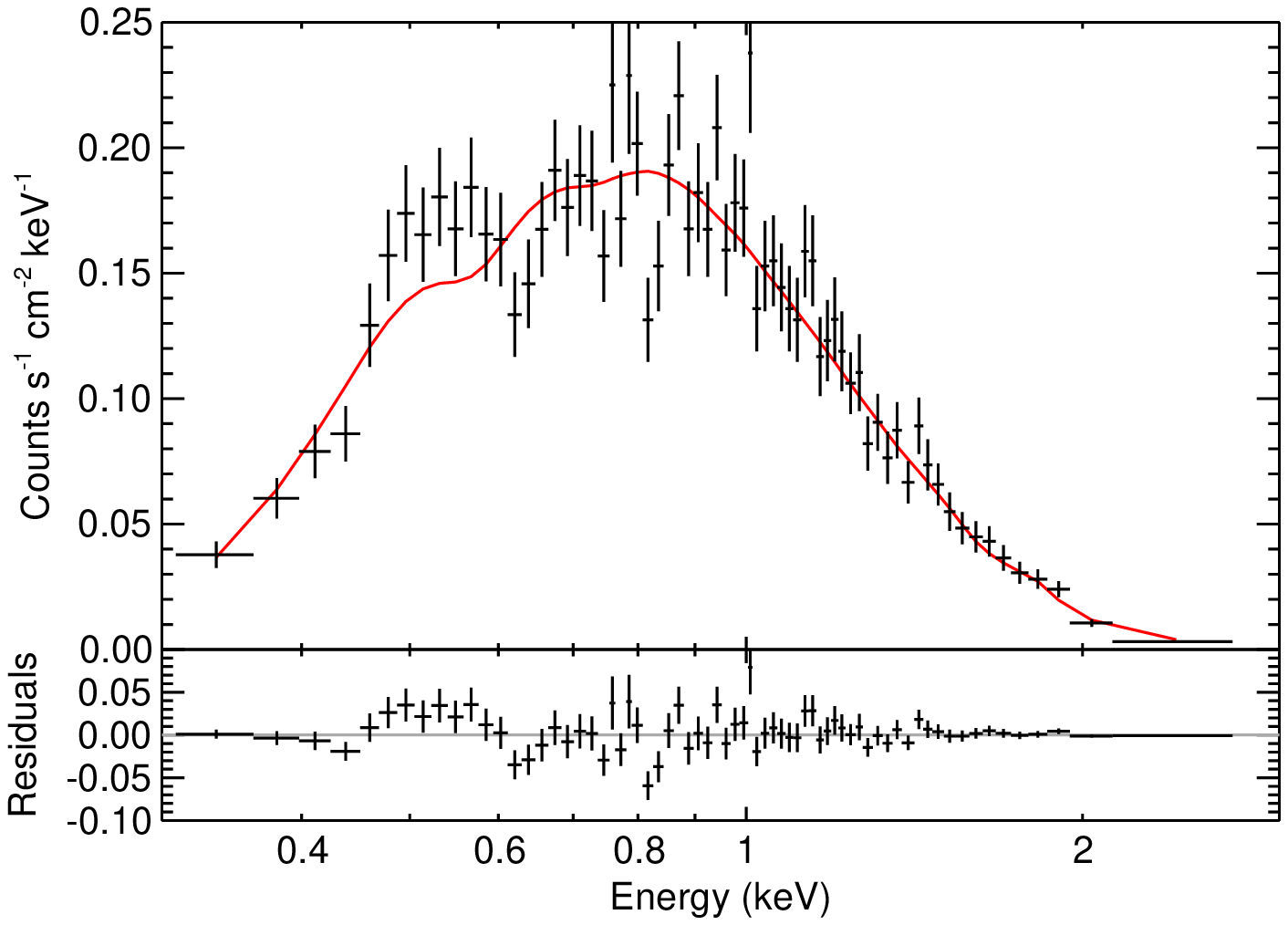}}
\bigskip
\caption[]{%
The \xray\ spectrum of \calvera\ (data), folded through the ACIS-S
response matrix and shown with our best-fit model (red line) and
residuals (lower panel).  The model consists of a non-magnetized
neutron star hydrogen atmosphere with gravitational mass $M=1.4\Msun$
and physical radius $R=12$\,km fixed, and having $kT_{\rm eff}=109\pm
1$\,eV, normalization $D_{\rm kpc}^{-2}=7.71^{+0.41}_{-0.38}\times
10^{-2}\,$, Galactic absorption $\nh=3.1\pm 0.9\times
10^{20}$\,\percmsq, and pile-up corrections (see text for details).
Data have been binned to a minimum of 50 source counts per bin.  The
best-fit spectrum is statistically acceptable ($\chisqrnu=1.305$ for
$\nu=67$, giving a probability of 4.9\%), but some systematic trends
in the fit residuals can be seen.  The most statistically significant
of these corresponds to a possible absorption feature at $E\approx
0.64$\,keV, or emission feature at $E\approx0.53$\,keV, as discussed
in the text.
\label{fig:nsa}}
\end{figure}

This fit can be seen in Fig.~\ref{fig:nsa}; the systematic trends in
the residuals below $E\approx 1$\,keV are suggestive and motivate
investigation of models with additional components.  


{\bf Neutron Star Hydrogen Atmosphere + Absorption Edge.}
Attempts to fit the spectrum with an unresolved absorption line at
$E\approx 0.65$\,keV did not achieve stable results.  However, fitting
the spectrum with a neutron star hydrogen atmosphere model plus
absorption edge ({\tt edge} and {\tt nsa} models in XSPEC), including
interstellar absorption and pile-up corrections, yields $\chisqrnu =
1.188$ ($\nu = 65$). This model has best-fit parameters: effective
temperature $kT_{\rm eff} = 109\pm 1$\,eV, equivalent neutral hydrogen
column density $N_H = 3.4\pm 0.8 \times 10^{20}$\,\percmsq, and
normalization $D_{\rm kpc}^{-2} = 0.079$.  The edge has best-fit
parameters of energy $E_{\rm edge} = 0.64^{+0.08}_{-0.06}$\,keV and
depth $\tau_{\rm edge} = 0.28\pm 0.12$ (Table~\ref{tab:params}),
corresponding to an equivalent width of ${\rm EW}\approx 70$\,eV. The
neutron star mass, radius, and magnetic field are fixed as before.

The addition of the edge feature results in a noticeable improvement
in the fit, $\Delta\chisqr = 10.1$, but we do not consider this
improvement significant for the reasons discussed below.


{\bf Neutron Star Hydrogen Atmosphere + Gaussian Emission Line.}
We also investigated an emission-line interpretation for the
low-energy residuals.  We find that an unresolved (width frozen at
zero) emission line ({\tt gaussian} model in XSPEC) added to the
hydrogen atmosphere model, including appropriate \nh\ absorption and
pile-up corrections, is statistically preferred to the absorption edge
fit. With $\chisqrnu=1.158$ ($\nu=65$), the model best-fit parameters
are $kT_{\rm eff} = 122\pm 3$\,eV, $N_H = 1.5\pm 1.0 \times
10^{20}$\,\percmsq, and normalization $D_{\rm kpc}^{-2} = 0.041$. The
line energy is $E_{\rm line} = 0.53\pm 0.02$\,keV with a normalization
of $2.48^{+1.4}_{-1.1} \times 10^{-5}$\,photon~cm$^{-2}$~s$^{-1}$,
corresponding to an equivalent width of ${\rm EW}\approx
28$\,eV. Other {\tt nsa} parameters remain fixed as before.

The addition of the emission line component provides an improvement of
$\Delta\chisqr = 12.2$, and is thus statistically preferred to the
absorption edge fit.  As we show below, however, it also falls short
of the 3$\sigma$ threshold that we would require to report detection.


\subsection{Monte Carlo Analyses}
\label{sub:obs:mc}

In order to investigate the statistical significance of the possible
low-energy absorption / emission features, and to estimate confidence
intervals on model parameters, we have carried out several Monte Carlo
analyses of the \calvera\ spectroscopic dataset.

To generate confidence intervals (Table~\ref{tab:params}), we perform
a bootstrap Monte Carlo analysis \citep{press}.  Drawing with
replacement from the photons making up our observed spectrum of
\calvera, we generate 10,000 bootstrap realizations of the spectrum,
fitting each realization with the spectral models described above.  We
perturb the start parameters for each fit randomly, according to
preliminary estimates of the parameter uncertainties, in order to
assure exploration of the $\chi^2$ landscape near minimum in every
case.  Typical final $\chi^2$ values for these automated fits are
higher than derived for the original data, as expected if the fits
fail to identify the global minimum for each dataset.  In the present
application, this can be expected to increase our estimates of
parameter uncertainties, yielding conservative estimates, and so is
not a significant concern.  Our quoted confidence intervals are
defined as the minimum-length intervals providing coverage of the
appropriate fraction of parameter values from these bootstrap trials.

To evaluate the statistical significance of the low-energy spectral
features, we generate a fake spectrum using the parameters of our
best-fit pure hydrogen atmosphere model and having the same number of
photons as the \calvera\ spectrum. We use this spectrum for a
bootstrap Monte Carlo analysis, generating $\approx$10,000 bootstrap
realizations of a \calvera-like spectrum that is now known to exhibit
no edge or line feature at low energies.  For each spectrum, we
perform a series of fits within XSPEC (after randomly perturbing the
parameter starting values, as previously).  First, we find the
best-fit hydrogen atmosphere spectrum.  Next, for a series of
prospective edge or line energies running from 0.3\,keV to 2.0\,keV,
in 0.1\,keV increments, we fit a hydrogen atmosphere + absorption edge
(or emission line) model.  The difference between the $\chi^2$ value
of the best-fit no-feature spectrum and the minimum $\chi^2$ value
from fits incorporating an absorption edge or emission line (of any
energy) is then recorded as the $\Delta\chi^2$ improvement for that
realization.  We treat the resulting distribution of $\Delta\chi^2$
values as the distribution of a test statistic, in order to evaluate
the probability of the null hypothesis that there is no absorption
edge (or emission line) present in the actual spectrum of \calvera.

The distribution of $\Delta\chi^2$ values from our bootstrap Monte
Carlo analysis for the emission line is shown in Fig.~\ref{fig:occam}.
Since $\Delta\chi^2 = 12.2$ for the actual \calvera\ spectrum, we find
the null hypothesis is disfavored at the $p=0.12$ or 88\%-confidence
level, i.e., the improvement in fit is not statistically
significant. The results for the absorption edge, which provides
$\Delta\chi^2 = 10.1$ for the actual data set, show that the null
hypothesis in this case is disfavored at the $p=0.19$ or
81\%-confidence level.

We note that these relatively low significance values are due in part
to our inability to specify the energy of the strongest absorption
feature or emission line a priori.


\section{Discussion}
\label{sec:discuss}

Our 30\,ks \chandra\ + ACIS-S observation of \calvera\ has served to
confirm the ICO nature of this source, by several means.  First, we
have confirmed the non-variable nature of \calvera\ on short
($10^{-4}\,{\rm Hz} \simlt \nu < 1.13$\,Hz) and long ($>$year)
time-scales.  Second, we have shown for the first time that a
blackbody fit is not adequate to explain the emergent spectrum of
\calvera\ over 0.3 to 7.0~keV; rather, a neutron star hydrogen
atmosphere model, with possible additional components, is required.
Third, the source retains its point-source appearance in this deepest
high-resolution \xray\ observation to-date.

Our observations do not resolve the fundamental conundrum of
interpretation for this source, first presented in \citet{rutledge08}.
Indeed, this conundrum is sharpened by our NS atmosphere fits, which
imply a distance of 3.6\,kpc (=$1/\sqrt{D_{\rm kpc}^{-2}}$; see Table
~\ref{tab:params}) for \calvera\ if its surface
\xray\ emission is nearly uniform, as in an INS scenario.  Our
measured column density, $\nh = 3.1\pm 0.9\times 10^{20}$\,\percmsq,
is consistent with the total Galactic column in the direction of
\calvera, estimated at $\nh = 2.65\times 10^{20}$\,\percmsq\ from
\ion{H}{1}\ maps \citep{kalberla05}.

The absence of slow \xray\ pulsations, to our limit of 8.0\% rms,
serves as a new and distinct aspect in which \calvera\ differs from
most (though not all) members of the INS population.  The frequency
range of the present search leaves the possibility of fast ($\nu\simgt
1$\,Hz) \xray\ pulsations untested; at the same time, the search
sensitivity is not sufficient to rule out low-amplitude, slow
pulsations such as those observed from \eighteen\ \citep{tiengo07}.
If \calvera\ is close to the Galactic plane, as expected given its
relatively high temperature, then it must exhibit non-uniform surface
emission which, in turn, is expected to produce \xray\ pulsations at
some level.  


\subsection{Spectrum}
\label{sub:discuss:spectrum}

Absorption features in the 0.1 to 1.0~keV energy range have been
observed, with varying degrees of significance, in a majority of INSs
at this point (see \citealt{haberl05}; \citealt{wkp+07}; and
references therein).  The interpretation of these features remains
unclear.  However, observation of multiple features in at least three
INSs, with energies lying in (or close to) harmonic relationships
(e.g., at energies 0.7 and 1.4\,keV for 1E~1207.4$-$5209;
\citealt{spz+02}), has led to their proposed interpretation as
cyclotron resonances of protons or electrons in the neutron star
atmosphere.  Alternatively, it has been argued that they may reflect
atomic features of highly-ionized atmospheric helium, oxygen, or neon
\citep{spz+02,mori07}.

If the possible absorption feature identified in our spectrum of
\calvera\ is real, then its properties are reminiscent of the known
INS absorption features, with the relatively higher energy $E\approx
640$\,eV suggesting a correspondingly stronger magnetic field in a
cyclotron interpretation.

Moreover, we note that photons with energies below 0.3~keV have been
excluded from our spectral analysis, as the ACIS response at these
energies is uncalibrated\footnote{See discussion at
  \url{http://asc.harvard.edu/cal/Acis/Cal\_prods/qeDeg}}.
Examination of the spatial distribution of lower-energy photons,
however, leaves no doubt that \calvera\ is detected as a point source
down to $E\approx 0.1$\,keV.  As such, it is possible that an
\xray\ spectral analysis of the broad range of 0.1--5.0\,keV emission
from \calvera\ at high signal-to-noise will reveal absorption features
at lower energies.  As a corollary, such an analysis might show our
present estimate of the column density to \calvera\ to have been
biased high by the presence of discrete (intrinsic) absorption near
0.3~keV.


\subsection{On Occam's Razor}
\label{sub:discuss:occam}

Detection of discrete features in \xray\ spectra requires adding at
least two parameters (the feature energy and its depth or
normalization) to any underlying spectral model.  The question of
whether the subsequent improvement in the fit statistic is
significant, given the number of added parameters, is properly treated
as an Occam's Razor problem (see, e.g., \citealt{magueijo07} and
references therein).

Quantitatively, Occam's Razor seeks to minimize the sum of the
information content of the data, given a theory, and the information
content of the theory itself. Accurately quantifying the latter, for
the general case of an arbitrary theory with any number of individual
parameters, has proven challenging; several distinct proposals have
been put forward in the statistical literature.  

\begin{figure}
\centerline{\includegraphics[width=7.2in]{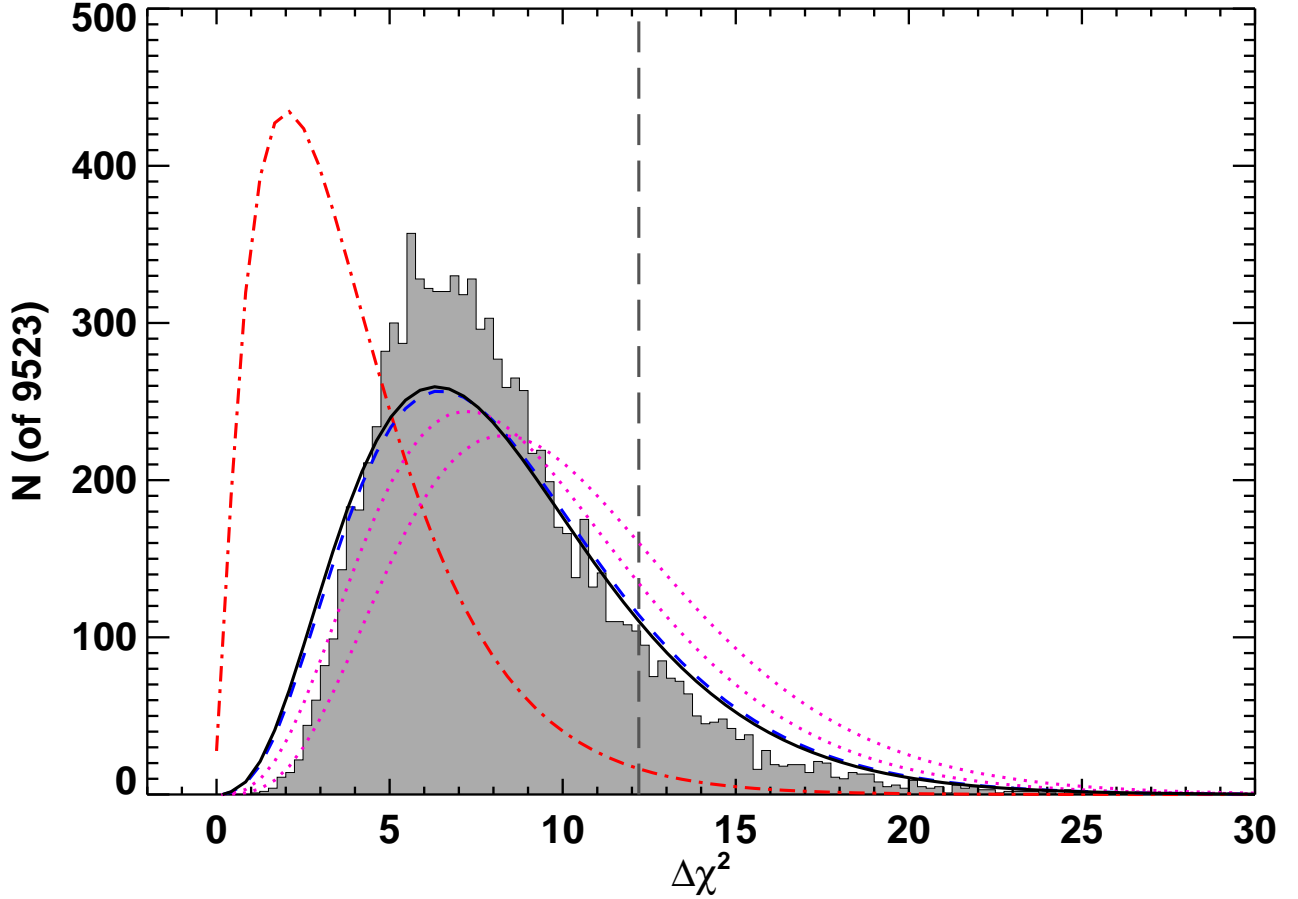}}
\bigskip
\caption[]{%
Distribution of $\Delta\chi^2$ values from our bootstrap Monte Carlo
analysis of the \calvera\ spectrum (histogram).  For each of 9,523
bootstrap realizations of a ``fake'' \calvera\ spectrum, known not to
exhibit an emission line, we calculate the difference in $\chi^2$
values between the best-fit line and no-line models.  We then use
this distribution as a test-statistic to evaluate the significance of
$\Delta\chi^2=12.2$, as observed for the actual \calvera\ spectrum
(vertical dashed line).  Our analysis suggests detection at
88\%-confidence and we conclude that the emission line is not a
statistically significant feature; an alternative absorption edge
interpretation is supported at 80\%-confidence in a similar analysis.  
Also shown on this plot are the $\chi^2$ distribution that best fits
the histogram of bootstrap values (black line, for $\nu=8.38$ degrees
of freedom), and the distributions predicted by several realizations
of Occam's Razor as applied to our analysis: the Akaike criterion (red
dash-dotted line, $\nu=4$); the Bayesian Information Criterion (blue
dashed line, $\nu=8.47$); and two realizations of the Sorkin criterion
(magenta dotted lines, $\nu=9.21$ and $\nu=10.27$).  See text for
detailed discussion.
\label{fig:occam}}
\end{figure}

In Fig.~\ref{fig:occam}, we present the predictions of three of these
theories and compare them to the results of our bootstrap Monte Carlo
analysis of the significance of an added emission line component in
the \calvera\ spectrum.  The distribution of $\Delta\chi^2$ values
between the models with and without emission line are predicted to
follow a $\chi^2$ distribution, for all theories; thus, the theory
predictions are realized as predictions for the number of degrees of
freedom ($\nu$) of this $\chi^2$ distribution.

According to the Akaike criterion \citep{akaike74}, we have added
$k=2$ free parameters to our model and expect a $\chi^2$ distribution
with $\nu=2k=4$, twice the number of added parameters.  According to
the ``Bayesian Information Criterion'' (BIC; \citealt{schwarz78}), on
the other hand, the number of data points should also be considered;
in this approach we expect $\nu=k\ln N\approx 8.47$, since in this
case $k=2$ and $N=69$.  Finally, under the Sorkin criterion
\citep{sorkin83}, we should estimate the additional information
content of the more complex model, evaluating the likely number of
distinguishable values (i.e., parameter range divided by parameter
uncertainty) for each added parameter $\ell_i$ and calculating $\nu=2
\Sigma_i \ln \ell_i$.  Applying the Sorkin criterion to any particular
analysis is thus less straightforward (see also \citealt{magueijo07});
in the case of our added emission line we estimate $9.21 < \nu <
10.27$.

As can be seen from Fig.~\ref{fig:occam}, the closest match to the
observed $\Delta\chi^2$ distribution has $\nu=8.38$; compared to the
various model predictions, this is consistent with the BIC estimate
$\nu=8.47$.  The Akaike criterion appears overgenerous for our case,
as it does not sufficiently penalize the more complex model for its
added parameters.  On the other hand, both of our Sorkin criterion
estimates -- meant to span the range of possibilities under this
approach -- seem overly conservative, predicting a significantly more
extended tail towards larger $\Delta\chi^2$ values.

Our bootstrap Monte Carlo results thus agree with the BIC prediction
as to the best-fit value of $\nu$; however, the distribution overall
does not follow a $\chisqr$ distribution (K-S test probability of
$p\sim 10^{-27}$).  We are satisfied with the bootstrap approach, and
present these results in detail to demonstrate the need for bootstrap
approaches as a validation and backstop to analytical prescriptions.

As an aside, we note that the $F$-test utterly fails to reproduce the
distribution of $\Delta\chi^2$ values from our bootstrap analysis.
Indeed, the $F$-test does not provide an appropriate statistical
metric for cases where the more complex model encompasses the simpler
model as a special case \citep{protassov02}.  Given the computer
resources presently available for Monte Carlo analyses, as well as the
well-developed literature on Occam's Razor, we hope that the $F$-test
is by now well and thoroughly discredited for these purposes.


\section{Conclusions}
\label{sec:conclude}

We have observed the isolated compact object \calvera\ for 30\,ks
with \chandra\ + ACIS-S in its 1/8-subarray mode and find no evidence
for large-amplitude ($>$8\% rms) slow pulsations ($\nu<1.13$\,Hz),
and no statistically-significant evidence of discrete spectral
features.  For the first time, we demonstrate that a simple blackbody
model is not an adequate fit to the emergent \xray\ spectrum of this
source, finding that a non-magnetized hydrogen atmosphere model (plus
Galactic absorption, after accounting for pile-up corrections)
provides a satisfactory fit to the \chandra\ data.

Our best-fit absorbing column density is consistent with the total
Galactic column measured from \ion{H}{1}\ surveys, and the nominal
distance estimate from our atmosphere fit is $d=3.6$\,kpc; thus, our
observations fail to resolve the conundrum of interpretation for
\calvera\ \citep{rutledge08}: as a full-surface emitter,
\calvera\ lies too far from the Galactic plane for its relatively hot
temperature unless it exhibits an extreme space velocity, $v_z \simgt
2000$\,\kmsec.  We identify a relatively bright nearby \xray\ source,
\objtwo, that might be used over a timescale of several years in an
\xray\ proper motion study to investigate this possibility.

Systematic trends in the residuals to our continuum spectral fits at
$E<1$\,keV suggest that low-energy emission or absorption features,
similar to those seen in other INS spectra, may be present in the
spectrum of \calvera.  Investigating two alternative approaches, we
first find evidence at the 80\%-confidence level for an absorption
edge at $E\approx 0.64$\,keV; alternatively, we find evidence at
88\%-confidence for an unresolved emission line at $E\approx
0.53$\,keV.  The physical explanation for either feature, if real,
remains unclear.  We note that photons from \calvera\ are detected
down to $E\approx 0.1$\,keV, so that a high signal-to-noise spectrum
extending to these energies should be able to efficiently identify and
characterize low-energy spectral features similar to those observed in
most INSs, if they are present.

Our evaluation of the statistical significance of the possible
low-energy spectral features uses a bootstrap Monte Carlo approach,
which we find satisfactorily addresses the difficulties involved in
Occam's Razor analyses.

Looking ahead, the most pressing need is to identify the pulsation
period for \calvera; apart from directly addressing the hypothesis
that \calvera\ is a nearby millisecond pulsar, pulse timing over an
extended period would allow a spin-down measurement and magnetic
field estimate for this object.  Should \calvera\ prove to be a fast
\xray\ pulsar, it would almost certainly be the subject of gravity-wave
searches using archived and future gravitational-wave observatory
datasets.

Observations using the \xmm\ EPIC-pn, for example, if they achieve
higher signal to noise than the present observation, would permit a
more sensitive search for pulsations, expand the frequency range up to
83\,Hz, confirm (independent of pile-up effects) that the
\xray\ spectrum is not consistent with a simple blackbody, and permit
a more detailed investigation of the marginally-significant absorption
or emission features we have here identified below 1~keV.


\acknowledgements

The authors acknowledge use of the online tools of NASA's High Energy
Astrophysics Science Archive Research Center, including WebPIMMS and
NH, and productive discussions and input from G.~G. Pavlov.
This work was supported by NASA Chandra General Observer funds under
award GO8-9075X. RER acknowledges support from the NSERC Discovery
Grant program.  


\bibliographystyle{apj_8}
\bibliography{complete}

\clearpage

\clearpage

\clearpage

\clearpage


\begin{deluxetable}{ll}
\tablewidth{12.2cm}
\tablecaption{Characteristics of \calvera\label{tab:params}}
\tablehead{
\colhead{Characteristic}  & 
\colhead{Value} 
}

\tabletypesize{\normalsize}

\startdata
Right Ascension (J2000)       & $14^{\rm h}\,12^{\rm m}\,55\fs 759$ \\
Declination (J2000)           & $+79\deg\,22\arcmin\,03\farcs 41$ \\
Uncertainty Ellipse           & 0.31\arcsec\ (R.A.) $\times$
                                0.25\arcsec\ (Dec.) \\
\hline
\multicolumn{2}{c}{Absorbed Blackbody\tablenotemark{a}} \\
\hline
\nh                        & 0 (limit) \\
\kteff                     & 229 eV \\
$(R_{\rm km}/D_{\rm 10\,kpc})^2$   & 26.6  \\  
Observed X-ray Flux (0.3--9.5 keV)  &  7.1\tee{-13} \cgsflux\ \\
\chisqrnu\ ($\nu$)          & 2.04 (67 dof) \\
\hline
\multicolumn{2}{c}{NS Hydrogen Atmosphere (NSA)\tablenotemark{b}}\\
\hline
\nh                        & 3.1\ud{0.9}{0.9} $\times$ 10$^{20}$ \percmsq \\
\kteff                     & 109\ud{1}{1} eV \\
$D_{\rm kpc}^{-2}$           & 7.71\ud{0.41}{0.38}\tee{-2} \\ 
Observed X-ray Flux (0.3--9.5 keV)  & 7.62\tee{-13} \cgsflux \\ 
\chisqrnu\ ($\nu$)          & 1.31 (67 dof) \\
\hline
\enddata

\tablenotetext{a}{The blackbody model does not provide a
  statistically acceptable fit and is listed for reference purposes
  only.}

\tablenotetext{b}{Additional parameters for {\tt nsa} models are set
  as follows: NS mass 1.4\,\Msun, radius 12\,\km, and magnetic field
  zero, corresponding to gravitational redshift $z_g = 0.235$.}

\tablecomments{All uncertainties are quoted at 90\%-confidence
  ($\pm$1.65$\sigma$ on a Gaussian distribution).} 

\end{deluxetable}

\clearpage


\begin{deluxetable}{ll}
\tablewidth{13cm}
\tablecaption{Possible Low-Energy Spectral Features in \calvera\label{tab:lowe}}
\tablehead{
\colhead{Characteristic}  & 
\colhead{Value} 
}

\tabletypesize{\normalsize}

\startdata
\hline
\multicolumn{2}{c}{NSA\tablenotemark{a}\ \ + 
                   Absorption Edge}\\
\hline
\nh                        & 3.4\ud{0.7}{0.8} $\times$ 10$^{20}$ \percmsq \\
\kteff                     & 109\ud{1}{1} eV \\
$D_{\rm kpc}^{-2}$           & 7.90\ud{0.39}{0.39}\tee{-2} \\
Edge Energy                & 0.64\ud{0.08}{0.06} keV \\
Edge Depth                 & 0.28\ud{0.12}{0.12} \\
Observed X-ray Flux (0.3--9.5 keV)  & 7.27\tee{-13} \cgsflux \\ 
\chisqrnu\ ($\nu$)          & 1.19 (65 dof) \\
\hline
\multicolumn{2}{c}{NSA\tablenotemark{a}\ \ + 
                   Emission Line}\\
\hline
\nh                        & 1.5\ud{1.0}{1.0} $\times 10^{20}$ \percmsq \\
\kteff                     & 122\ud{3}{3} eV \\
$D_{\rm kpc}^{-2}$           & 4.08\ud{0.59}{0.54}\tee{-2} \\
Line Energy                & 0.53\ud{0.02}{0.02} keV \\
Line Normalization         & 2.48\ud{1.4}{1.1}\tee{-5} photons~cm$^{-2}$~s$^{-1}$ \\
Observed X-ray Flux (0.3--9.5 keV)  & 7.41\tee{-13} \cgsflux \\
\chisqrnu\ ($\nu$)         & 1.16 (65 dof) \\
\hline
\enddata

\tablenotetext{a}{A neutron star hydrogen atmosphere ({\tt nsa})
  is adopted as our continuum model in both cases. Additional model
  parameters are set as follows: NS mass 1.4\,\Msun, radius 12\,\km,
  and magnetic field zero.  The corresponding gravitational redshift is
  $z_g = 0.235$.} 

\tablecomments{The absorption edge and emission line feature are not
  considered statistically significant; see text for details.
  Uncertainties are quoted at 90\%-confidence ($\pm$1.65$\sigma$ on a
  Gaussian distribution) and are provided for reference purposes
  only.}

\end{deluxetable}


\end{document}